# Rootlets-based registration to the spinal cord PAM50 template

**Authors:** Sandrine Bédard[1], Jan Valošek[1,2,3,4], Valeria Oliva[5,6], Kenneth A. Weber II[5], and Julien Cohen-Adad[1,2,7,8]

**AFFILIATIONS:**

1. NeuroPoly Lab, Institute of Biomedical Engineering, Polytechnique Montreal, Montreal, QC, Canada
2. Mila - Quebec AI Institute, Montreal, QC, Canada
3. Department of Neurosurgery, Faculty of Medicine and Dentistry, Palacký University Olomouc, Olomouc, Czechia
4. Department of Neurology, Faculty of Medicine and Dentistry, Palacký University Olomouc, Olomouc, Czechia
5. Neuromuscular Insight Lab, Division of Pain Medicine, Stanford School of Medicine, Palo Alto, CA, United States.
6. Center for Behavioral Sciences and Mental Health, Italian National Institute of Health, Rome, Italy
7. Functional Neuroimaging Unit, CRIUGM, Université de Montréal, Montréal, QC, Canada
8. Centre de Recherche du CHU Sainte-Justine, Université de Montréal, Montréal, QC, Canada

**ORCID:**
Sandrine Bédard - 0000-0001-9859-1133
Jan Valošek - 0000-0002-7398-4990
Valeria Oliva - 0000-0001-7849-191X
Kenneth A Weber II 0000-0002-0916-9174
Julien Cohen-Adad - 0000-0003-3662-9532

# Abstract

Spinal cord functional MRI studies require precise localization of spinal levels for reliable voxelwise group analyses. Traditional template-based registration of the spinal cord uses intervertebral discs for alignment. However, substantial anatomical variability across individuals exists between vertebral and spinal levels. This study proposes a novel registration approach that leverages spinal nerve rootlets to improve alignment accuracy and reproducibility across individuals.

We developed a registration method leveraging dorsal cervical rootlets segmentation and aligning them non-linearly with the PAM50 spinal cord template. Validation was performed on a multi-subject, multi-site dataset (n=267, 44 sites) and a multi-subject dataset with various neck positions (n=10, 3 sessions). We further validated the method on task-based functional MRI (n=23) to compare group-level activation maps using rootlet-based registration to traditional disc-based methods.

Rootlet-based registration showed superior alignment across individuals compared to the traditional disc-based method. Notably, rootlet positions were more stable across neck positions. Group-level analysis of task-based functional MRI using rootlet-based increased Z scores and activation cluster size compared to disc-based registration (number of active voxels from 3292 to 7978).

Rootlet-based registration enhances both inter- and intra-subject anatomical alignment and yields better spatial normalization for group-level fMRI analyses. Our findings highlight the potential of rootlet-based registration to improve the precision and reliability of spinal cord neuroimaging group analysis.

# 1. Introduction

Substantial anatomical variability in the central nervous system across populations, sexes, and age groups has led to the development of population-based templates of the brain (Evans et al., 2005; Fonov et al., 2011) and spinal cord (De Leener et al., 2018). Templates allow the standardization of individual magnetic resonance imaging (MRI) images into a common space, enabling group-level and atlas-based analyses and improving reproducibility across studies, as shown for microstructural mapping in the spinal cord (Lévy et al., 2015). A key application of template-based analysis is spatial normalization, which is essential for group-level analyses, particularly in functional MRI (fMRI) (Kinany et al., 2020; Oliva et al., 2025; Weber et al., 2018) and diffusion MRI studies (Pisharady et al., 2020; Vallotton et al., 2021). Additionally, spatial normalization facilitates the creation of population-based probabilistic maps of pathologies, such as spatiotemporal distribution of multiple sclerosis lesions in the spinal cord (Eden et al., 2019).

In spinal cord imaging, the PAM50 template (De Leener et al., 2018) is widely used for anatomical alignment, and includes a probabilistic atlas of white and gray matter (De Leener et al., 2018; Lévy et al., 2015). Template registration typically involves straightening the spinal cord and aligning vertebral levels—identified by intervertebral discs—with those in the template. However, vertebral levels are defined on bony landmarks, whereas spinal levels are defined neuroanatomically by the entry zones of spinal nerve rootlets (Cadotte et al., 2015; Diaz and Morales, 2016). Importantly, there is substantial inter-subject variability in the correspondence between vertebral and spinal levels (Cadotte et al., 2015; Diaz and Morales, 2016; Mendez et al., 2021), particularly along the caudal axis, where rootlets become increasingly angulated (Frostell et al., 2016; Valošek et al., 2024). This discrepancy is further influenced by neck position (flexion, neutral, or extension) (Bédard et al., 2023; Cadotte et al., 2015).

The differences between vertebral and spinal levels present a challenge for spinal fMRI, where accurate localization of spinal levels is critical for reliable group-level analyses (Kinany et al., 2020). Conventional approaches rely on disc-based registration to the template and infer spinal levels from the population average, which can obscure individual-level differences (Frostell et al., 2016; Kinany et al., 2022). A more precise approach would involve identifying spinal levels directly in the subject's native space. This is now feasible through the automatic segmentation of nerve rootlets (Valošek et al., 2024) as part of the Spinal Cord Toolbox (SCT) (De Leener et al., 2017a). The C2-C8 dorsal spinal nerve rootlets can be segmented on T2-weighted (T2w) MRI scans and used to estimate spinal levels.

In this work, we propose a novel spatial normalization method that incorporates spinal nerve rootlets information into the registration process. We hypothesized that using the rootlet segmentations for registration would improve inter-subject alignment accuracy by accounting for individual neuroanatomical landmarks. To validate this approach, we compared the rootlet-based approach against the current state-of-the-art disc-based approach using two open-access datasets of T2w images. We also applied both methods to a task-based fMRI study to evaluate the impact of rootlet-based registration on spatial normalization and functional activation maps.

# 2. Methods

## 2.1. Rootlet-based registration to the PAM50 template

An overview of the rootlet-based registration pipeline is shown in **Figure 1**. The method is implemented in SCT v7.0 (De Leener et al., 2017a) via the command `sct_register_to_template -lrootlet`. It requires two inputs: a binary spinal cord segmentation and a level-specific segmentation of the dorsal cervical rootlets from C2 to C8 (2: C2 rootlet, 3: C3 rootlet, etc.). Both the spinal cord and rootlets segmentation can be generated automatically by SCT's deep learning models via `sct_deepseg spinalcord` and `rootlets_t2` commands, respectively (Valošek et al., 2024). Since the rootlets segmentation model is currently limited to T2w images, users working with other contrasts can alternatively provide manual segmentations. Note that ongoing work is focused on extending the automatic rootlets segmentation method to other contrasts and levels, as well as ventral rootlets (see Limitations and Perspectives).

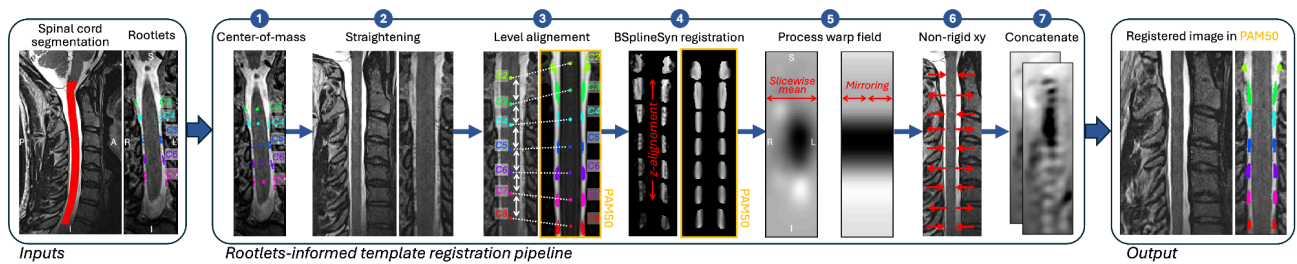

*Figure 1. Rootlet-based template registration.* *Input segmentations of the spinal cord and rootlets undergo the following steps: (1) center-of-mass calculation per level, (2) spinal cord straightening, (3) non-linear alignment of levels to the PAM50 template, (4) BSplineSyn registration along the superior-inferior axis using the rootlet segmentation as a mask on both subject and PAM50 T2w images, (5) slice-wise averaging of warping field for right-left symmetry, (6) xy-plane (axial) scaling adjustment, and (7) concatenation of warping fields for complete transformation to the template.*

The main steps of the pipeline are detailed below:

1. **Center-of-mass calculation:** The center-of-mass of each segmented rootlets level is computed for the subject native image and for the PAM50 template (**Figure 1**, step 1).

2. **Straightening:** Spinal cord is straightened using Non-Uniform Rational B-Splines (NURBS) (De Leener et al., 2017b). The spinal cord segmentation is used to compute the centerline that is then used to perform straightening. The straightening is required as the PAM50 template is a straightened spinal cord image.

3. **Initial alignment of rootlets levels:** The extracted center-of-mass for each rootlet level is aligned non-linearly with the ones in the PAM50 template (**Figure 1**, steps 3).

4. **Non-linear registration in superior-inferior axis:** The rootlets segmentation is warped to the straightened image (using a warping field from the previous steps, 2 and 3), dilated by 3 voxels and used to mask the straightened subject image. Similarly, the rootlets segmentation in the PAM50 space (already straightened) is used to mask the PAM50 T2w image. Then, a non-linear registration (BSplineSyn, (Tustison et al., 2021)) is performed along the superior-inferior on the masked, straightened image, to refine the alignment of the rootlets with the PAM50 (**Figure 1**, step 4). The parameters of this registration (step 4) can be customized using the flag `-param type=rootlet`.

5. **Processing of warping field:** The resulting warping field from the superior-inferior alignment (step 4) is averaged slice-wise to ensure right-left symmetry (**Figure 1**, step 5). The mean is calculated excluding zero values.

6. **xy-plane scaling:** A non-rigid registration in the xy-plane (axial) is applied to adjust the cord size to match the PAM50 template (**Figure 1**, step 6) (De Leener et al., 2018).

7. **Warping field concatenation:** The forward and backward transformations to the PAM50 template are generated by concatenating the warping fields from steps 2, 3, 5 et 6 (**Figure 1**, step 7).

## 2.2. Validation

We compared the proposed rootlet-based registration to the traditional disc-based registration on three datasets: a multi-subject, multi-site dataset of T2w images (Section 2.2.1), a dataset of T2w images from 10 healthy participants scanned across three neck positions (Section 2.2.2), and a task-based spinal cord fMRI dataset (Section 2.2.3).

### 2.2.1. Inter-subject rootlets alignment

#### 2.2.1.1. Data & participants

We used 3T T2w 0.8×0.8×0.8 mm$^3$ images covering the entire cervical spine of 267 healthy participants from the open-access Spine Generic multi-subject dataset (Cohen-Adad et al., 2021). The dataset includes participants aged 19–56 years (50% female) scanned across 44 centers worldwide, using three MRI vendors (Siemens, GE, Philips).

#### 2.2.1.2. Analysis pipeline

For each T2w image, the spinal cord was segmented using `sct_deepseg spinalcord`, cervical dorsal rootlets (C2-C8) were obtained using `sct_deepseg rootlets_t2` (Valošek et al., 2024), and vertebral levels were identified using `sct_label_verterbrae` (Ullmann et al., 2014). When available, existing spinal cord segmentations and intervertebral disc labels from the Spine Generic dataset were used (Cohen-Adad et al., 2021). Images were then registered to the template using both rootlet-based (described in section 2.1) and disc-based template registration (De Leener et al., 2018). Results were visually inspected using SCT's quality control feature `sct_qc` (Valošek and Cohen-Adad, 2024). Participants were excluded if rootlets segmentation was incomplete (missing an entire level) or substantially under-segmented, for example, due to mild compression, canal narrowing, or lower cerebrospinal fluid contrast. The final cohort consisted of 226 participants.

The processing time for both registration methods was recorded. To assess anatomical alignment, we averaged the images of each participant in the template space for each method (rootlets vs. discs) and computed the overlap of warped rootlets in the superior-inferior axis to the template space with the PAM50 rootlets using both registration methods. The overlap was computed for each participant as the following:

$$Overlap_i = \frac{length_{overlap}}{(slice_{inf} - slice_{sup})}$$

With $length_{overlap}$ as the number of slices along the superior-inferior axis in which the subject's rootlet segmentation overlaps with the PAM50 rootlets for level $i$, $slice_{inf}$ and $slice_{sup}$ as the most inferior and most superior slices of the participant's rootlet segmentation for that level $i$, respectively.

### 2.2.1.3. Morphometric analysis

To evaluate the impact of rootlet-based vs. disc-based registration on the anatomical alignment across individuals, we analyzed spinal cord morphometry in the template space. Specifically, we examined whether rootlet-based registration improved cervical enlargement alignment across individuals. An overview of the pipeline is presented in **Figure S1**.

For this analysis, the final xy-plane (axial) scaling step (**Figure 1**, step 6) was omitted in both registration methods to preserve native spinal cord morphology while maintaining alignment along the superior-inferior axis. Spinal cord segmentation was repeated in the template space using `sct_deepseg spinalcord` to minimize interpolation errors.

Spinal cord cross-sectional area (CSA) was computed slice-wise using `sct_process_segmentation` and normalized over 20 slices (10 mm) centered at C2–C3 intervertebral disc level—just above the cervical enlargement—to reduce inter-subject variability (Bédard and Cohen-Adad, 2022; Cohen-Adad et al., 2021). To improve visualization, the normalized CSA was smoothed using a 45-slice (22.5 mm) window, and local maxima were identified to determine the center of the cervical enlargement.

Participants with radiological signs of mild spinal cord compression (Valošek et al., 2024a) were excluded from the CSA analysis, reducing the final cohort to 176 participants.

### 2.2.2. Intra-subject rootlets alignment

To evaluate the effect of the different neck positions on the registration performance, we compared both rootlets and discs registration methods using the analysis pipeline described in [section 2.2.1.2](#) on a dataset of 10 healthy participants scanned across three neck positions (flexion, neutral and extension) with 3T 0.6×0.6×0.6 mm$^3$ T2w images (Bédard et al., 2023; Bédard and Cohen-Adad, 2023).

### 2.2.3. Validation on task-based fMRI analysis

#### 2.2.3.1. Data & participants

Task-based fMRI data were used to assess the impact of rootlet-based vs. disc-based registration on the group-level activation maps. Data from 28 healthy participants were collected at Stanford Lucas Center for Imaging on a 3T GE SIGNA Premier scanner with a 21-channel neurovascular coil (Oliva et al., 2025). Briefly, anatomical T2w scans (0.8×0.8×0.8 mm$^3$, 3D turbo spin-echo) and functional images (2D spatially selective reduced FOV pulse, 15 slices, 1.25×1.25×5.00 mm$^3$) with the field-of-view centered at C5-C6 intervertebral discs were acquired. Participants performed a right-handed sequential finger-tapping task at 1 Hz. Visual cues were provided with *Eprime* software (Version 2.0, Psychology Software Tools, Pittsburgh, PA). The task started with a 15-second rest block, followed by 10 trials of 15 seconds of the task intertwined with 15-second resting periods.

#### 2.2.3.2. Methods

Data processing is detailed in Oliva et al. (2025) and included motion correction, physiological noise filtering, and spatial normalization using both rootlet-based and disc-based registration. A Gaussian smoothing kernel (2×2×5 mm) was applied. Participants with missing spinal levels in the rootlets segmentation were excluded, reducing the number of participants from 28 to 23. Subject-level activity maps were generated and analyzed at the group-level with a fixed-effects model with cluster correction (voxel-wise threshold Z > 3.10, and cluster threshold p = 0.05 ) to examine activation patterns (Oliva et al., 2025).

# 3. Results

## 3.1. Computational efficiency

Processing of 267 participants was distributed across 15 CPU cores, with each core handling one participant on a 64-core CPU cluster (8x Intel Xeon E7-4809 2.10GHz). The rootlet-based registration method took an average of 783.3 ± 46.74s (13.05 min) per participant compared to 747.4 ± 57.4s (12.45 min) per participant for the disc-based method.

## 3.2. Inter-subject rootlets alignment

**Figure 2** presents the mean image of 226 participants from the Spine Generic dataset (Cohen-Adad et al., 2021), after registration to the PAM50 template using either rootlet-based or disc-based registration. The rootlet-based method provides sharper delineation of both dorsal and ventral rootlets (red arrows in **Figure 2**), whereas the disc-based method results in blurred rootlet structures due to increased misalignment.

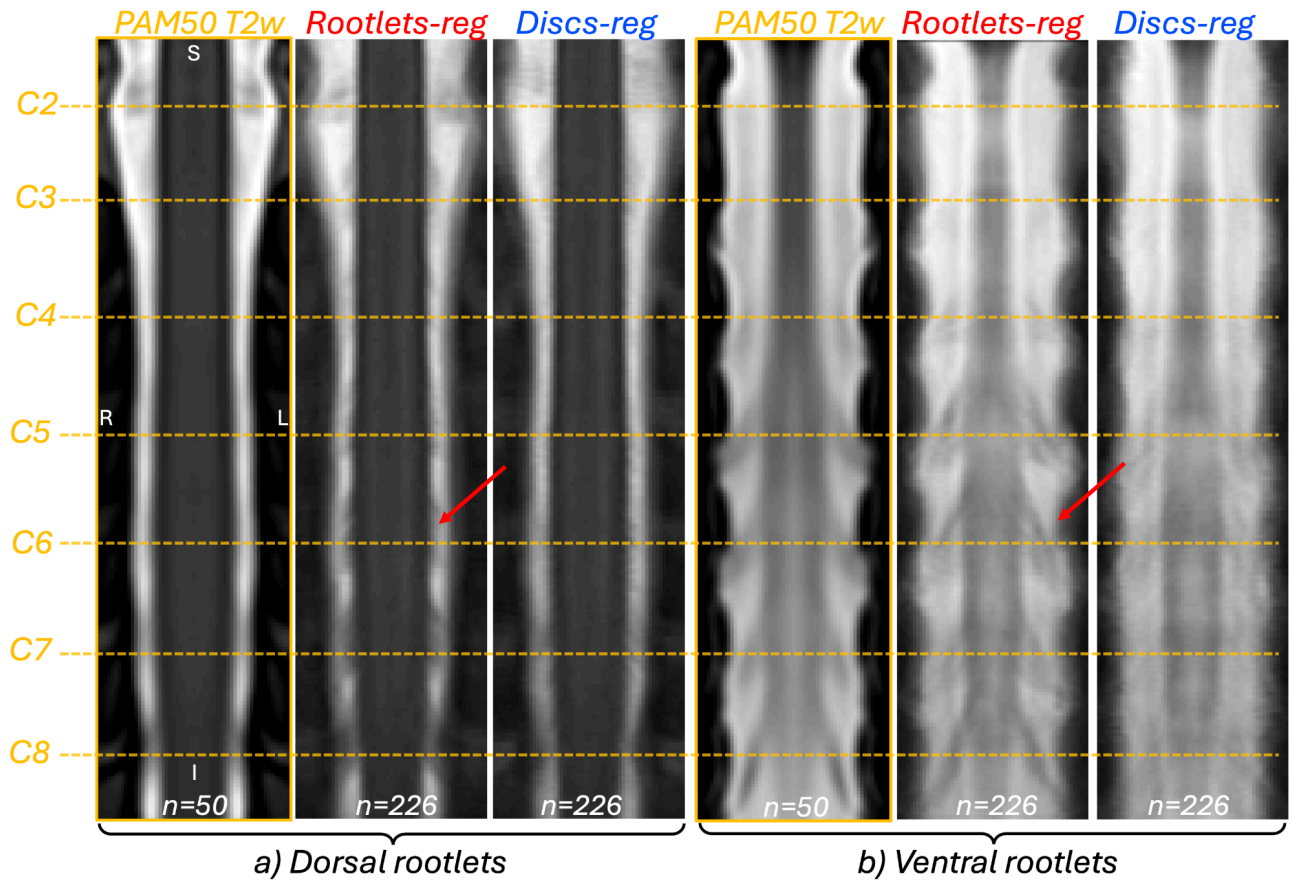

a) Dorsal rootlets    b) Ventral rootlets

***Figure 2.*** *Mean T2w image (n=226) registered to the template using rootlet-based (red) or disc-based (blue) registration. The PAM50 T2w image (n=50) is also shown (orange). The center of spinal levels C2 to C8 as identified in the PAM50 template is marked by an orange dashed line. **a)** Example coronal slice (y=67) showing dorsal rootlets. **b)** Example coronal slice (y=78) showing ventral rootlets in PAM50 template space. Red arrows indicate an example of a good delineation of spinal rootlets. Across 226 participants, the averaged rootlets appear sharper with the rootlet-based registration than with the disc-based method, demonstrating superior alignment.*

**Figure 3** shows the average coverage of dorsal spinal rootlet segmentations along the superior–inferior axis in PAM50 space across 226 participants. Rootlet-based registration yielded mean rootlet positions that more closely aligned with the PAM50 template compared to the disc-based method, which exhibited greater variability. The standard deviation (STD) of rootlet coverage was also lower with the rootlet-based method, reflecting greater consistency across participants. In contrast, the disc-based approach showed higher STD values, of the upper and lower bounds of rootlet coverage, as illustrated by the shaded areas in **Figure 3**. This variability was especially pronounced at more caudal spinal levels (e.g., C8) for the disc-based method. Across all spinal levels, the mean overlap between participant rootlets and PAM50 rootlets in the superior-inferior axis was 89.62 ± 10.11% for the rootlet-based method versus 64.94 ± 21.09% for the disc-based method (with 100% describing a perfect overlap with the PAM50 rootlets).

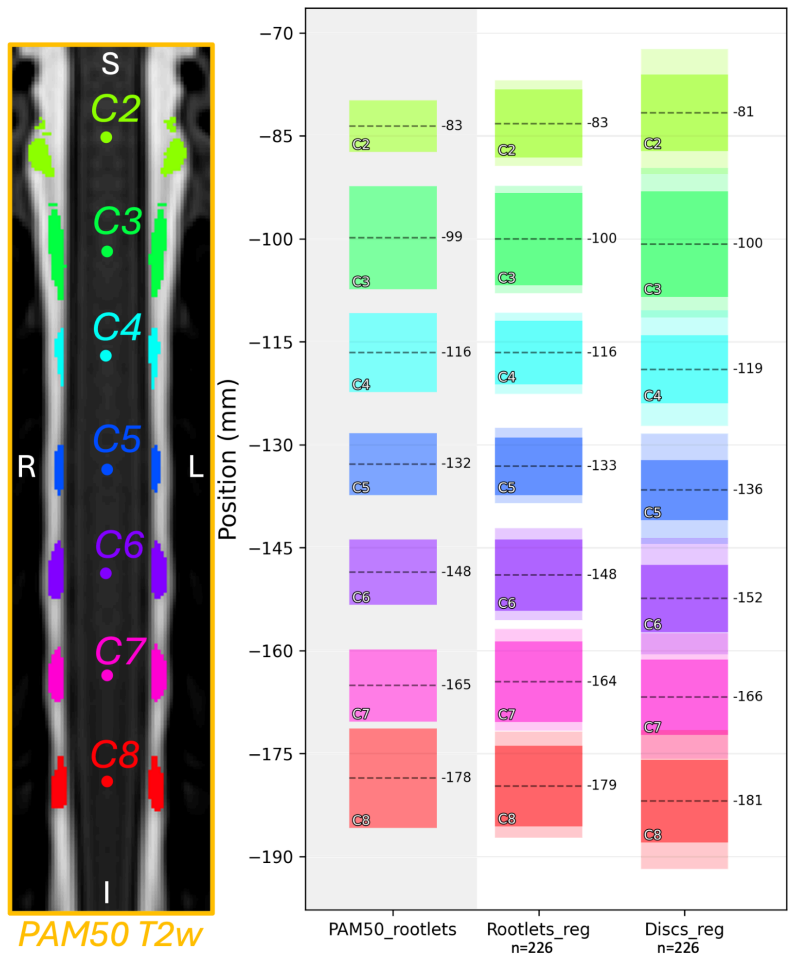

*Figure 3.* Dorsal rootlets coverage in the superior-inferior axis in the PAM50 space from n=226 participants, registered using rootlet-based (Rootlets_reg) or disc-based (Discs_reg) methods. Each box plot shows the mean center position (dashed line) and extent of the rootlet coverage. The shaded area represents the STD of the inferior and superior bound of the coverage across participants. The PAM50 rootlets are shown on the left panel (shaded). Note that in the PAM50 space, 1 mm corresponds to 2 slices (0.5×0.5×0.5 mm$^3$ resolution).

### 3.2.1. Morphometric analysis

To assess whether cervical enlargement alignment across participants improves with rootlet-based registration, we analyzed spinal cord CSA in the template space for rootlets (**Figure 4a**) and discs (**Figure 4b**) registration methods, normalized at C2-C3 intervertebral discs in template space.

The local maxima of each participant's CSA curves (indicating the cervical enlargement) are shown as dots in **Figure 4**. The mean position of the cervical enlargement in the superior-inferior axis expressed as z-coordinate in the PAM50 space was 843.09 ± 9.87 for the rootlet-based registration and 837.03 ± 11.76 for the disc-based registration. The rootlet-based method results in greater alignment along the superior-inferior axis, as evidenced by the more consistent positioning of cervical enlargements across participants (**Figure 4c**) and lower STD.

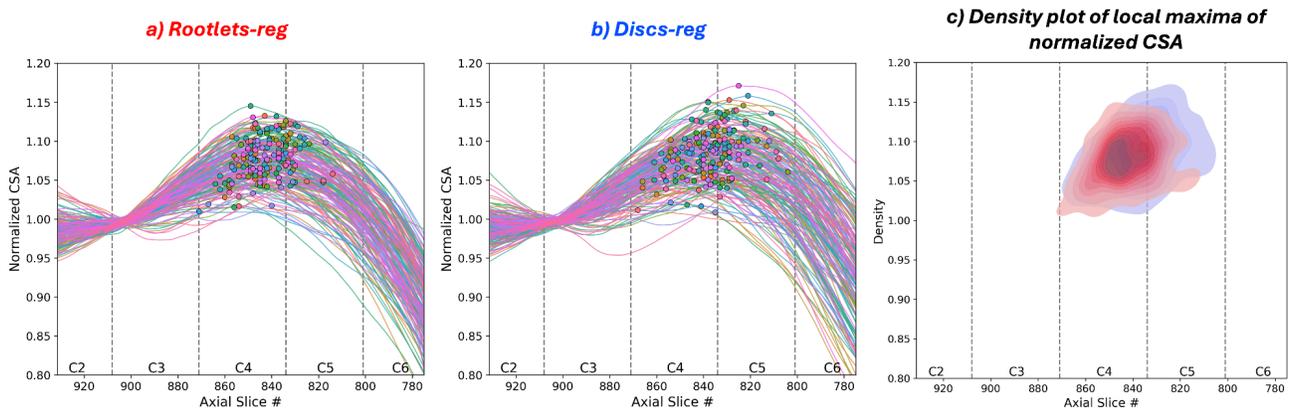

**Figure 4.** *Cross-sectional area (CSA) in the PAM50 space obtained using **(a)** rootlet-based registration and **(b)** disc-based registration methods in n=176 participants. CSA was normalized at C2-C3 disc, smoothed with a 45-slice window, and plotted with local maxima (single dot). **(c)** Density plot of the local maxima for both rootlet-based (red) and disc-based (blue) methods. Vertebral levels and axial slices in the PAM50 space are identified, one line represents one participant. Participants with mild compression were excluded.*

### 3.3. Intra-subject rootlets alignment

**Figure 5** shows registration results for a participant scanned in three different neck positions (flexion, neutral, and extension). **Figure S2** provides an example for another participant.

With rootlet-based registration, the dorsal rootlet positions remained stable in the superior-inferior axis across neck positions, showing consistent alignment with the PAM50 template. In contrast, disc-based registration led to greater misalignment (yellow brackets in **Figure 5**), particularly in the extension position.

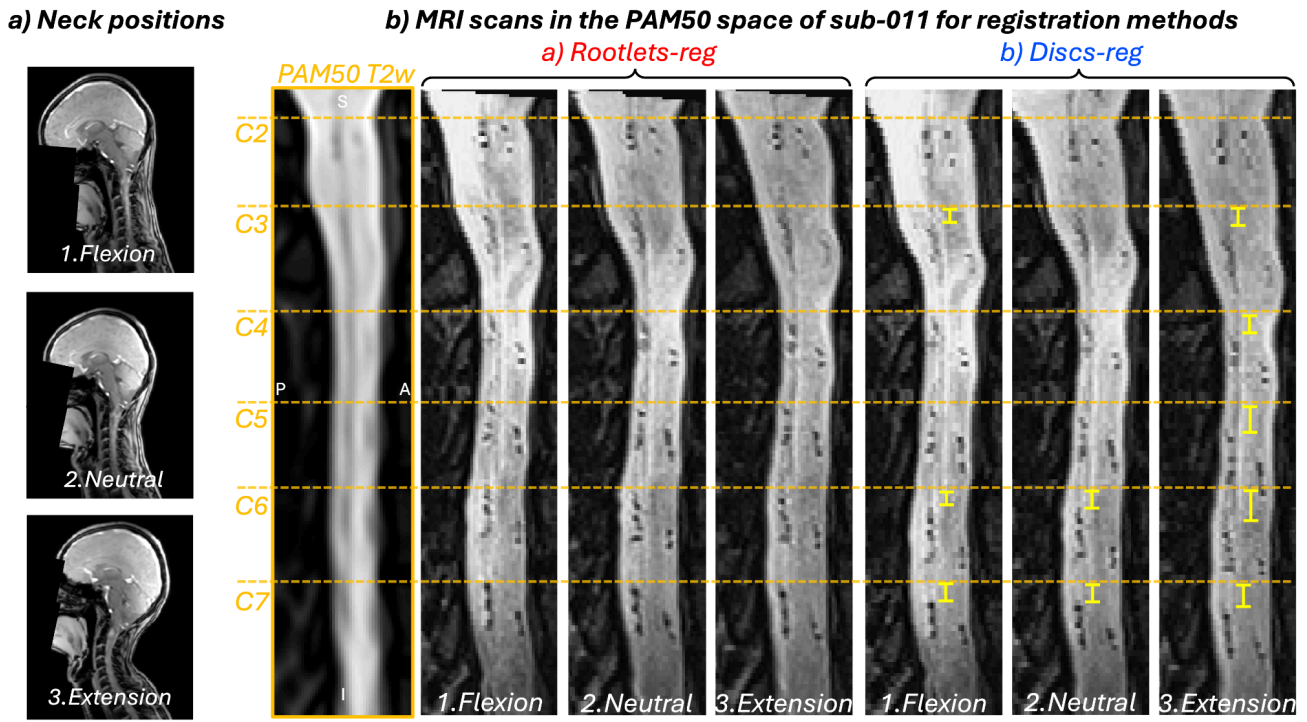

***Figure 5.*** *(a) Illustration of different neck positions: flexion (1), neutral (2) and extension (3). (b) Registration results for participant sub-011, showing spinal cord alignment using rootlet-based (red) and disc-based (blue) registration in all three neck positions. The PAM50 sagittal slice (x=86) is displayed for reference. The orange dashed lines mark the top of spinal levels C2-C7. Yellow brackets highlight the misalignment of dorsal rootlets in the disc-based registration method compared to the PAM50.*

### 3.4. Validation on task-based fMRI analysis

The mean registered anatomical image across 23 healthy participants is shown in **Figure S3** for both registration methods. Consistent with the results from [section 3.2](#), rootlet-based registration allows clearer delineation of both dorsal and ventral rootlets compared to the disc-based approach.

Group level activation maps from the task-based fMRI study (n=23) are shown in **Figure 6**, comparing rootlet-based and disc-based registration methods. We observe that the rootlet-based method resulted in larger activation clusters compared to disc-based registration. The density plot of active voxels and Z score shows an increase in activation intensity for the rootlet-based method compared to the disc-based method, and is more localized at C6 and C7 spinal levels**.** Mean Z score for disc-based registration was 4.04 ± 0.77 compared to 4.43 ± 1.09. The number of active voxels for rootlet-based registration was 7978 compared to 3292 for disc-based registration.

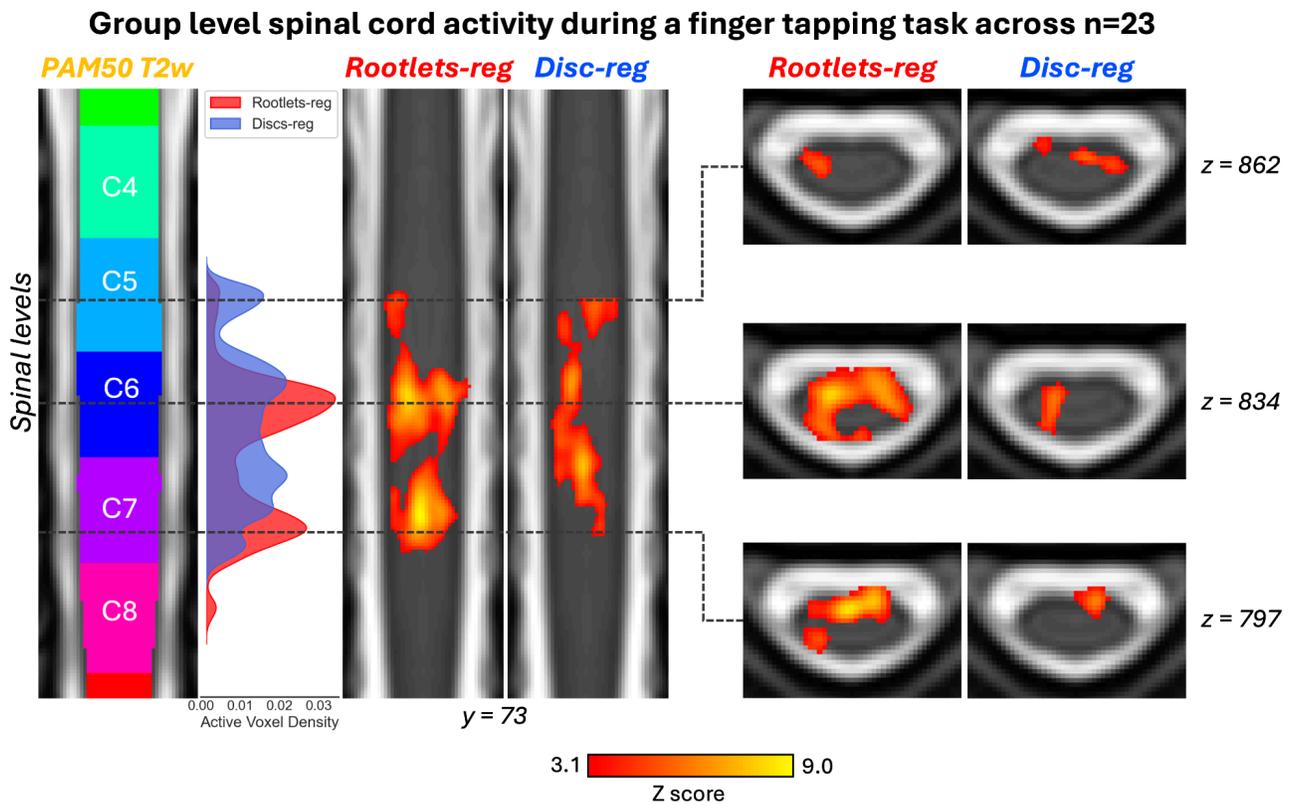

*Figure 6.* Group-level activation maps during a sequential finger-tapping task (n=23) using rootlet-based (red) and disc-based (blue) registration for spatial normalization, in the PAM50 space. *(a)* Coronal slice (y=73) showing vertebral levels, density plot of active voxels and activation maps. *(b)* Activation maps on 3 axial slices. The activation maps were generated using a fixed-effects model, voxel-wise thresholded at Z score > 3.10, and cluster-corrected for family-wise error (FWE) at $p < 0.05$. The PAM50 T2w template is used as the background, with spinal levels indicated. The comparison reveals larger activation clusters and higher z-scores with rootlet-based registration compared to the disc-based method.

# 4. Discussion

This study introduced a novel method to improve template registration of spinal cord MRI data that leverages spinal nerve rootlets. Compared to traditional intervertebral disc-based registration, our rootlet-based approach yielded more accurate anatomical alignment on a large cohort of healthy participants (n=226), across 10 subjects with different neck positions, and resulted in increased cluster size and Z scores in the context of task-based spinal cord fMRI (n=23).

## 4.1. Inter-subject rootlets alignment

Rootlet-based registration produced more consistent alignment of both dorsal and ventral cervical rootlets in the template space across individuals relative to disc-based registration. Compared to the PAM50 template, constructed using images from 50 healthy individuals, the rootlet-based method achieved clearer delineation of the rootlets—particularly the dorsal rootlets—even when applied to a larger cohort (n=226 vs n=50 for the PAM50 template). Note that the PAM50 template was constructed based on intervertebral discs alignment and also included smoothing (De Leener et al., 2018).

When comparing the rootlets coverage in the superior-inferior axis across individuals in the PAM50 space between registration methods (**Figure 3**), we observed larger errors for disc-based registration at more caudal levels (e.g. C8). As previously reported, differences between spinal levels and vertebral levels increased for lower levels (Cadotte et al., 2015; Valošek et al., 2024). With the cervical enlargement located at spinal levels C5–C6, where disc-spinal discrepancies are smaller, the localization of the enlargement between methods was comparable (**Figure 4**) with slightly better alignment for the disc-based method as indicated by lower STD. Importantly, rootlet-based registration did not compromise the alignment of cervical cord morphology, confirming its potential for preserving anatomical integrity while achieving improved alignment. Additionally, the position and size of the cervical enlargement has substantial variability across healthy individuals, thus improving the superior-inferior axis alignment may not mitigate the entire inter-subject variability (Frostell et al., 2016; Nunès et al., 2023).

## 4.2. Intra-subject alignment across neck positions

Rootlet-based registration also showed advantages for aligning spinal cord images within the same participants across sessions with different neck positions, confirming a previous study by Bédard et al. (2023). In contrast, disc-based registration exhibited poor intra-subject alignment, especially in extreme positions such as extension. While such postural differences during MRI scans are more exaggerated than typical inter-session variation in clinical settings, this experiment highlights the limitations of disc-based methods in accommodating even minor variations in head and neck positioning—an important consideration for improving test-retest reliability in spinal cord fMRI (Dabbagh et al., 2024; Kowalczyk et al., 2024).

### 4.3. Impact on task-based fMRI

Group-level task-based fMRI analysis demonstrated that rootlet-based registration yielded larger activation clusters and higher Z scores. These results suggest better anatomical correspondence across participants. It is important to note that smoothing applied during fMRI preprocessing (5 mm Gaussian kernel in superior-inferior axis) may attenuate the effects of fine-grained improvements in alignment. Future studies should assess whether rootlet-based registration enhances sensitivity in other contexts, such as resting-state connectivity or test-retest scenarios. Multi-site validation will also be essential to ensure reproducibility across imaging protocols and hardware.

### 4.4. Limitations and Perspectives

The current rootlet segmentation method is restricted to T2w images covering spinal levels C2-C8 (Valošek et al., 2024). Ongoing work aims to adapt the segmentation method to other contrasts, ventral rootlets, and lower spinal levels, although spinal levels might be more difficult to distinguish due to the limited resolution of in-vivo MRI scans. Methods such as those described in Liang et al. (2025) could provide insights into extending rootlet segmentation to lower spinal levels. Also, severe spinal cord compression or spinal canal narrowing may impair rootlets visibility, reducing registration reliability in pathological populations.

The method relies on center-of-mass localization of rootlets rather than their true entry zone into the spinal cord, which is difficult to identify in standard MRI due to the lack of contrast between rootlets and surrounding pia mater. Note that a second step using directly the rootlets to refine this z-alignment is used to minimize the error associated with the center-of-mass computation. In the context of application to fMRI data where the acquisition includes relatively thick slices (5 mm) in addition to spatial smoothing, these minor errors are unlikely to substantially impact group-level functional analyses.

While the overall processing time was comparable between the two methods, correcting segmentation errors in rootlets can be more time-consuming, particularly when rootlets are poorly visualized due to image quality or anatomical variability. In contrast, manual correction of disc labels is typically simpler and faster. Incorporating ventral rootlets may improve robustness, particularly in compressed cords where dorsal structures are displaced. Finally, the absence of normative MRI-based studies on rootlet positioning in healthy individuals limits our ability to define a ground truth for evaluation.

A potential application of this work is the development of a new spinal cord template aligning spinal rootlets, which could offer more precise spatial normalization than the current PAM50 template, originally constructed using disc-based alignment (De Leener et al., 2018). Another possible avenue is evaluating the rootlets-based registration method in patients with spinal cord compression, where the rootlets-based alignment might facilitate the registration of PAM50 atlas to DWI space. Finally, the proposed rootlet-based registration can be applied to fMRI, especially for improving group-level fMRI outcomes, enhancing intra-subject reliability, and enabling more sensitive detection of functional changes in both healthy and clinical populations. By validating this method across multiple sites and imaging paradigms, we hope to establish it as a standard for high-precision spinal cord image analysis.

## 5. Conclusion

Rootlet-based registration offers improved anatomical alignment over traditional disc-based methods for cervical spinal cord MRI. This approach enhances both inter- and intra-subject alignment and yields better spatial normalization for group-level fMRI analyses. Our findings highlight the potential of rootlet-based registration methods to advance the precision and reliability of spinal cord neuroimaging analysis.

## 6. Acknowledgements

We thank Nick Guenther and Mathieu Guay-Paquet for their assistance with the management of the datasets, Joshua Newton for his contributions in helping us implement the algorithm to SCT. We also thank Christine S. W. Law, Merve Kaptan, Dario Pfyffer, Brett Chy, Susanna Aufrichtig, Nazrawit Berhe, Gary H. Glover and Sean Mackey for their contribution in data acquisition of the task-based fMRI data.

## 7. Funding

SB is supported by the Natural Sciences and Engineering Research Council of Canada, NSERC, Canada Graduate Scholarships — Doctoral program. JV received funding from the European Union's Horizon Europe research and innovation programme under the Marie Skłodowska-Curie grant agreement No 101107932. VO is supported by the Italian National Institute of Health (Starting Grant for Young Researchers CUP I85E23001090005).

Funded by the Canada Research Chair in Quantitative Magnetic Resonance Imaging [CRC-2020-00179], the Canadian Institute of Health Research [PJT-190258], the Canada Foundation for Innovation [32454, 34824], the Fonds de Recherche du Québec - Santé [322736, 324636], the Natural Sciences and Engineering Research Council of Canada [RGPIN-2019-07244], the Canada First Research Excellence Fund (IVADO and TransMedTech), the Courtois NeuroMod project, the Quebec BioImaging Network [5886, 35450], INSPIRED (Spinal Research, UK; Wings for Life, Austria; Craig H. Neilsen Foundation, USA), Mila - Tech Transfer Funding Program. This study was supported by grants from the National Institute of Neurological Disorders and Stroke (grant numbers K23NS104211, L30NS108301, R01NS128478, and R01NS133305). The content is solely the responsibility of the authors and does not necessarily represent the official views of the National Institutes of Health.

## 8. Ethics

All datasets used in this study complied with all relevant ethical regulations.

## 9. Data/Code Availability

The data used in this study come from open-access datasets and can be accessed for Spine Generic at https://github.com/spine-generic/data-multi-subject/releases/tag/r20250310, and for *Spinal Cord Head Positions* dataset at https://openneuro.org/datasets/ds004507/versions/1.1.1. Task-based fMRI data will be made available on open-neuro.

All analysis code is openly accessible on GitHub at https://github.com/sct-pipeline/rootlets-informed-reg2template/releases/tag/r20250428. The rootlet-based registration method is available in Spinal Cord Toolbox (https://github.com/spinalcordtoolbox/spinalcordtoolbox/releases/tag/7.0) v7.0 and higher via `sct_register_to_template -lrootlet`.

## 10. Author Contribution

S.B: Data Curation, Formal Analysis, Investigation, Methodology, Visualization, and Writing (original draft, review & editing).

J.V: Data Curation, Formal Analysis, Investigation, Methodology, Visualization, and Writing (original draft, review & editing).

V.O.: Data Curation, writing (review and editing).

K.W: conceptualization, data curation, funding acquisition, investigation, methodology, supervision, writing (review and editing).

J.C.A: conceptualization, data curation, funding acquisition, investigation, methodology, supervision, writing (review and editing).

## 11. Competing Interests

The authors declared no potential conflicts of interest with respect to the research, authorship, and/or publication of this article.

# 13. Supplementary Material

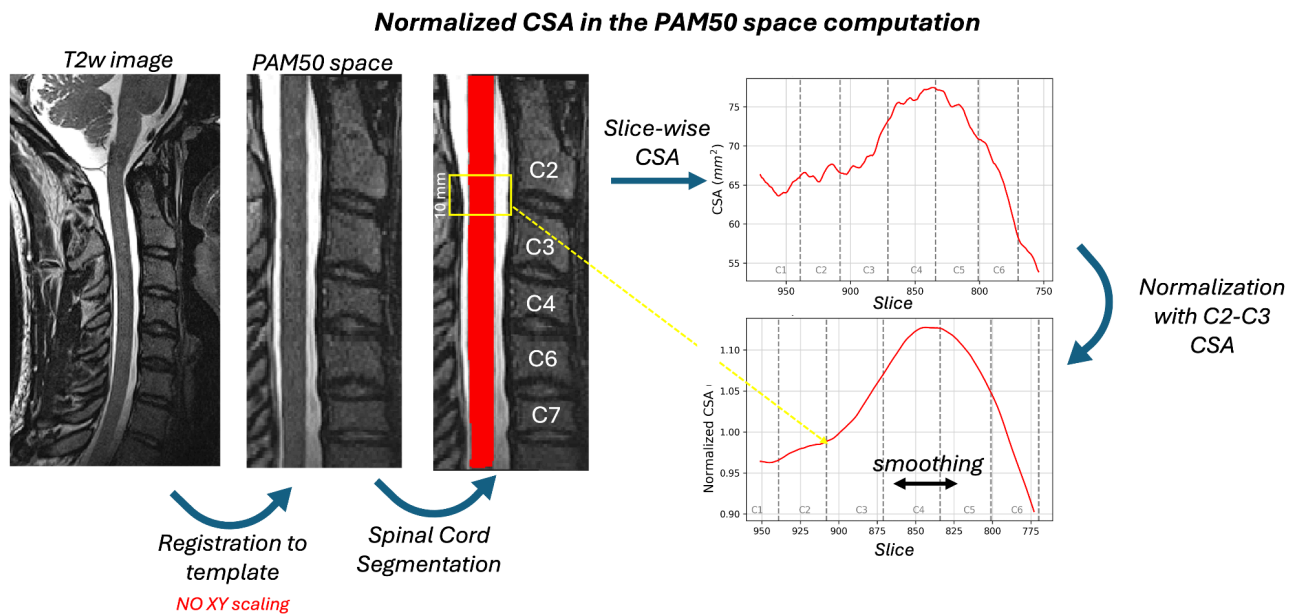

*Normalized CSA in the PAM50 space computation*

*Figure S1. Pipeline to analyse spinal cord morphometry.* First, the T2w image is registered to the PAM50 template using either rootlet-based or disc-based registration without the axial scaling of the spinal cord shape (xy-plane) to avoid changing the spinal cord morphology. Then, the spinal cord was segmented and the spinal cord cross-sectional area (CSA) was computed for each slice. CSA was normalized to the over 20 slices (10 mm) centered at C2–C3 intervertebral disc level and smoothed using a 45-slice window.

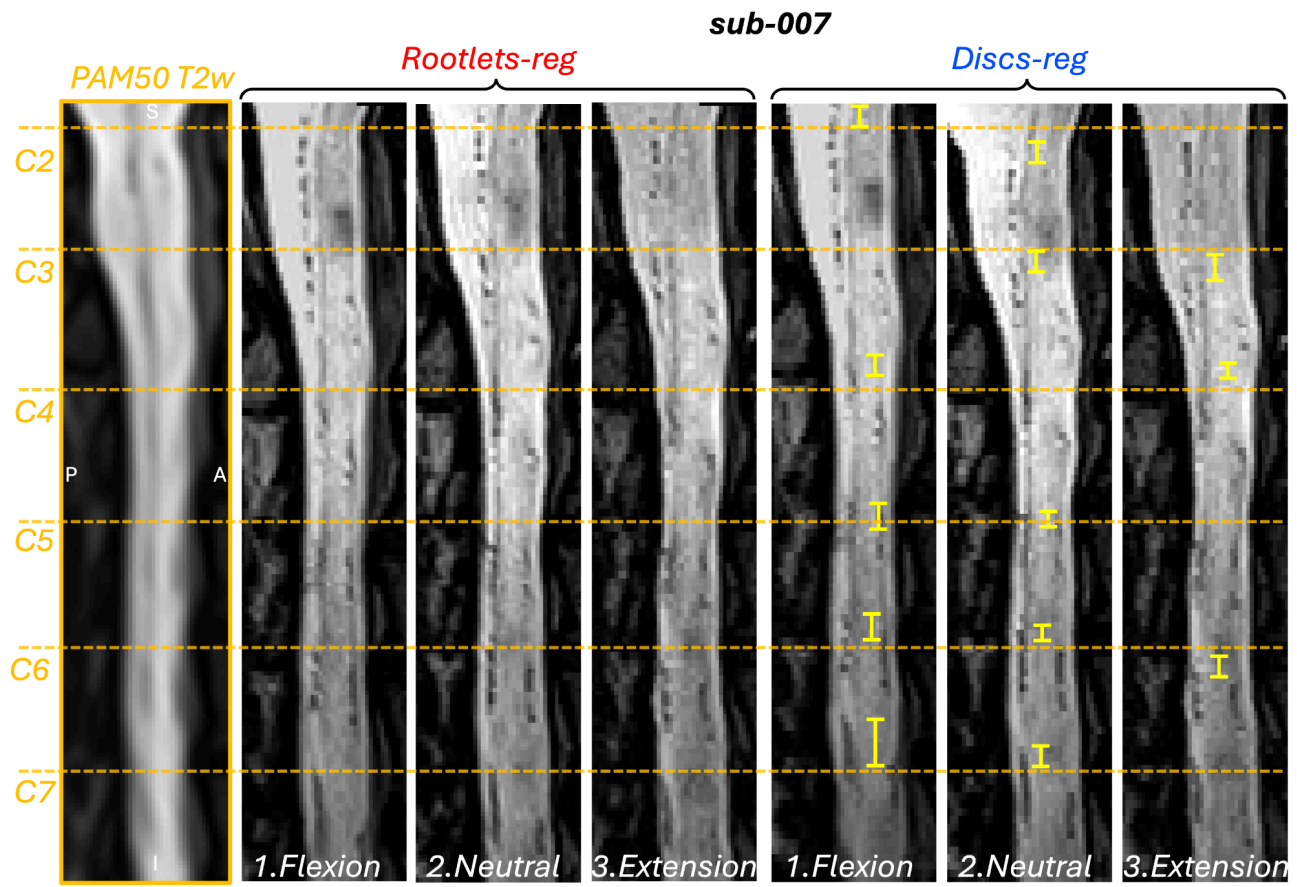

**Figure S2.** *Registration results for participant sub-007, showing spinal cord alignment using rootlet-based (red) and disc-based (blue) registration in all three neck positions. The PAM50 sagittal slice (x=84) is displayed for reference. The orange dashed lines mark the top of spinal levels C2-C7. Yellow brackets highlight misalignment of dorsal rootlets in the disc-based registration method compared to the PAM50.*

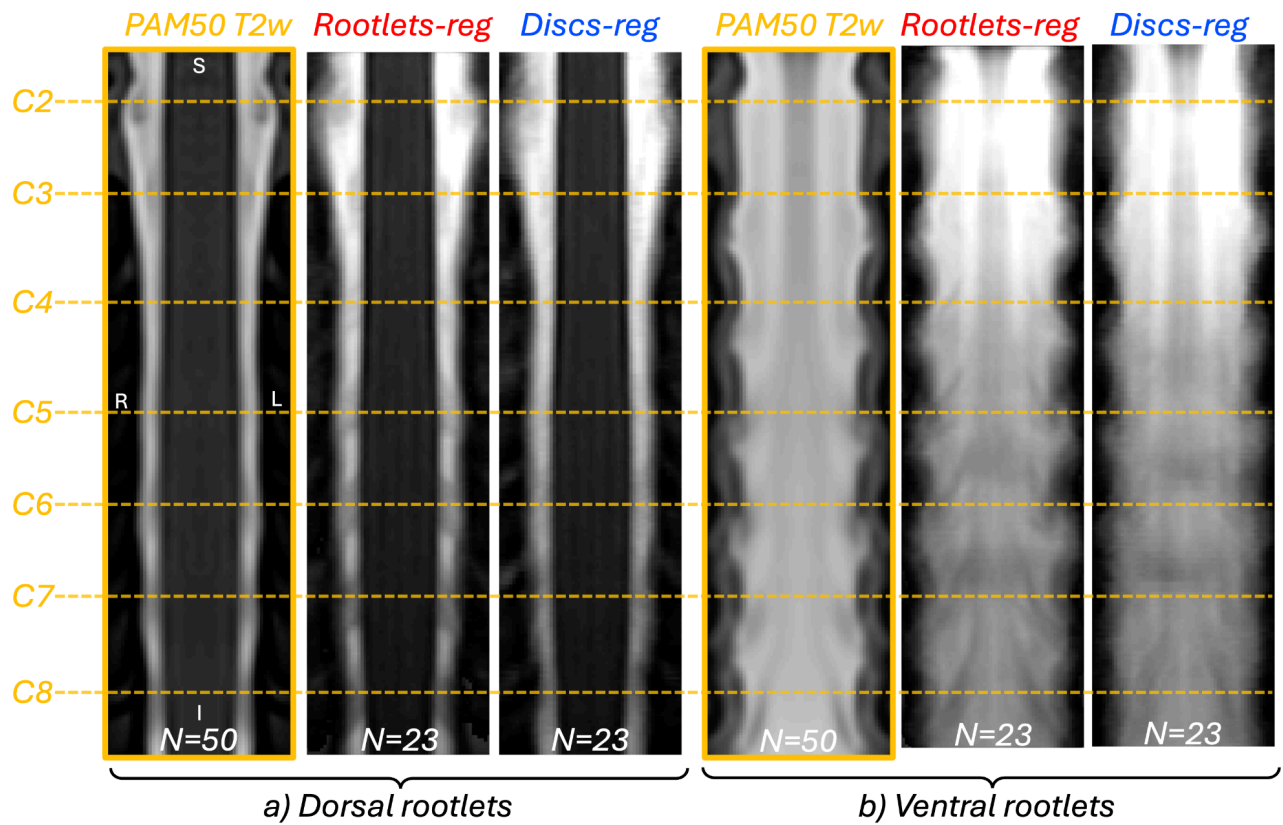

**Figure S3.** *Mean T2w image (n=23) registered to the template using rootlet-based (red) or disc-based (blue) registration. The PAM50 T2w image (n=50) is also shown (orange). The center of spinal levels C2 to C8 in the PAM50 template is marked by an orange dashed line. **a)** Example coronal slice (y=67) showing dorsal rootlets. **b)** Example coronal slice (y=78) showing ventral rootlets in PAM50 template space. Red arrows indicate an example of a good delineation of spinal rootlets.*

23